%% file: main.tex
\documentclass[runningheads]{llncs}

\usepackage{graphics}
\usepackage{csquotes}
\usepackage{todonotes}
\usepackage{tikz}
\usepackage{hyperref}
\usepackage{comment}

\usepackage[position=b]{subcaption}
\usepackage{caption}
\usepackage{graphicx}

\usepackage[maxbibnames=99]{biblatex}
\begin{document}
\title{Enabling Versatile Privacy Interfaces Using Machine-Readable Transparency Information}

\titlerunning{Enabling Versatile Privacy Interfaces}
\author{Elias Grünewald\inst{1}\orcidID{0000-0001-9076-9240} \and \\
Johannes M. Halkenhäußer\inst{1}\orcidID{0000-0001-6538-7531} \and \\
Nicola Leschke\inst{1}\orcidID{0000-0003-0657-602X} \and \\
Johanna Washington\inst{2}\orcidID{0000-0002-5693-6575} \and \\
Cristina Paupini\inst{3}\orcidID{0000-0003-4139-6331} \and \\
Frank Pallas\inst{1}\orcidID{0000-0002-5543-0265}}

\authorrunning{Grünewald et al.}
\institute{Technische Universität Berlin, Information Systems Engineering, Germany\\
\email{\{eg, j.halkenhaeusser, nl, fp\}@ise.tu-berlin.de} \and
iRights.Lab, Berlin, Germany\\ \email{j.washington@irights-lab.de} \and
Oslo Metropolitan University, Norway\\ \email{cristina.paupini@oslomet.no}}
\maketitle              %

\begin{abstract}
Transparency regarding the processing of personal data in online services is a necessary precondition for informed decisions on whether or not to share personal data. In this paper, we argue that privacy interfaces shall incorporate the context of display, personal preferences, and individual competences of data subjects following the principles of universal design and usable privacy. Doing so requires -- among others -- to consciously decouple the provision of transparency information from their ultimate presentation.
To this end, we provide a general model of how transparency information can be provided from a data controller to data subjects, effectively leveraging machine-readable transparency information and facilitating versatile presentation interfaces. We contribute two actual implementations of said model: 1) a GDPR-aligned privacy dashboard and 2) a chatbot and virtual voice assistant enabled by conversational AI. We evaluate our model and implementations with a user study and find that these approaches provide effective and time-efficient transparency. Consequently, we illustrate how transparency can be enhanced using machine-readable transparency information and how data controllers can meet respective regulatory obligations.
\end{abstract}


\usetikzlibrary{positioning, shadows, shapes.geometric, shapes.symbols, patterns}
\newcommand*\pointer[1]{\tikz[anchor=2mm]{\node[shape=circle,fill=black, text=white,scale=0.5] (char) {\textbf{#1}};}}

\input{content}

\section*{Acknowledgements}

First, we thank Flora Muscinelli and Michael Gebauer for their work on the development and operations of our prototypes. Moreover, we thank the DaSKITA\footnote{\url{http://tu.berlin/ise/daskita}} project team for their valuable support. We also thank Maximilian~von~Grafenstein and Julie Heumüller for fruitful discussions and their input on the privacy dashboard. In addition, we thank the team of the CheckMyVA project for jointly recruiting the participants of the user study.

This research was funded within the project TOUCAN\footnote{\url{https://tu.berlin/ise/toucan}}, supported under grant no. 01IS17052 by funds of the German Federal Ministry of Education and Research (BMBF) under the Software Campus 2.0 (TU Berlin) program.

\renewcommand{\contentsname}{References}
\renewcommand{\bibname}{References}

\printbibliography
\end{document}

%% file: content.tex
\section{Introduction}

Transparency is a core principle for protecting data subjects' privacy worldwide and %
one of the leitmotifs of the usable privacy discipline \cite{reuter2022usablePrivacy}: data subjects shall be well-informed about the consequences and potential risks of their interactions with systems that process personal data relating to them. %
Although regulations, such as the European General Data Protection Regulation (GDPR) or the California Consumer Privacy Act (CCPA), oblige data controllers to provide transparent information about their data processing practices, traditional ways for doing so -- as in written privacy policies -- fail to convey relevant details \cite{schaub2015}. As a consequence, poorly informed data subjects may, e.g., provide consent to dubious or even malevolent processing of sensitive data.

Over the last years, an emerging field of usable privacy technologies can be observed \cite{reuter2022usablePrivacy}. In particular, comprehensive privacy iconographies have been or are being developed. %
Amongst them are privacy-focused nutrition labels \cite{kelley2009nutrition}, icons indicating notice \& and choice mechanisms \cite{schaub2017designing}, icons embedded in browser extensions \cite{cranor2006user}, and many more \cite{hedbom2008survey, gruenewaldPallas2021daten}. Still, their evaluation shows that data subjects need recurring contexts, previous experience, and learning to correctly understand their contents \cite{rossi2020}.
Especially, the interpretation of abstract legal terms and concepts remains challenging. %

For privacy visualizations and the related human-computer interactions, we identify at least three major requirements. First, these approaches need to fulfill legal requirements, i.e., provide sufficient expressiveness to depict the transparency information obligations from the applicable regulatory framework (e.g., the GDPR or CCPA). Second, the information needs %
to be provided in a machine-readable format to process large-scale sharing infrastructures. Art. 12(7)~GDPR also explicitly states this obligation in the context of privacy icons for accessibility reasons. Third, the designs are to be evaluated for their effectiveness, as shown for the \enquote{Do not sell my information}) opt-out mechanism established by the CCPA \cite{togglesDollar2021}.

Similarly, data protection authorities (DPAs) observe 
more and more tactics of %
\enquote{dark patterns}, which describe malformed privacy-related interfaces, that make use of, e.g., overloading, misleading information, or decontextualization \cite{edpb2022}. Given that data subjects \textit{(i)} behave differently depending on the context they are using a service in \cite{nissenbaum2004privacy,tsai2009impact, windl2022privacy}, \textit{(ii)} have heterogeneous personal preferences, and \textit{(iii)} bring a different level of competence relating to the processing of personal data, there is an urgent need for adaptive privacy interfaces. We, therefore, argue to avail oneself of the Universal Design approach which aims for \enquote{services to be usable by all people to the greatest extent possible} \cite{un2006crpd}. More specifically, it takes into account the multiplicity of ways in which people access and navigate society \cite{imrie2012universalism, ostroff2011universal}. Consequently, service providers can guarantee the usability of their product while protecting equality and non-discrimination at the same time \cite{giannoumis2019conceptualizing}. 

The intensive scientific discourse notwithstanding, 
widespread adoption of user-friendly and at the same time legally aligned mechanisms for providing transparency information is still missing. Instead, the above-mentioned \enquote{dark patterns} dominate the experience of the modern web, and even promising privacy icon sets have not been widely implemented. A major barrier to actual adoption lies in the fact 
that %
proposed solutions typically focus on overly specific use case scenarios and are therefore not transferable across contexts with reasonable costs and efforts. %
To address this gap, %
we herein provide the following contributions:

\begin{enumerate}
    \item A general reference model for exchanging transparency information between a data controller and a data subject based on Universal Design principles, that can be used to implement and evaluate transparency enhancing technologies.
    \item Two open-source implementations of this model to demonstrate its real-world applicability to enable versatile (i.e. context-, preference-, and competence-adaptive) privacy information guided by Universal Design principles. More specifically, these %
    process machine-readable transparency information in the form of
    \begin{enumerate}
        \item an extendable and layered privacy dashboard for transparency based on participatory user workshops, and
        \item a privacy preference-focused chatbot and voice assistant powered by conversational artificial intelligence. 
    \end{enumerate} %
    \item A preliminary evaluation of the proposed model, as materialized in said implementations, including a user study involving 19 data subjects. %
\end{enumerate}

\noindent Hence, we organize this paper as follows:
In section~\ref{sec:related-work}, we provide some background and related work.
Section~\ref{sec:model} describes a general reference model that can be used to provide transparency information.
Afterward, we describe two implementations (in section~\ref{sec:implementation}) of the reference model, namely a layered privacy dashboard in section~\ref{sec:privacy-icons} and an interactive privacy chatbot and voice assistant in section~\ref{sec:chatbot}.
We evaluate our model in section~\ref{sec:evaluation}.
Finally, section~\ref{sec:outlook} gives an overview of future work and concludes.

\section{Background and Related Work} \label{sec:related-work}

Transparency requirements from privacy regulations such as the GDPR constitute the starting point for our considerations.
According to Art.~12 ff.~GDPR, for example, data subjects must be informed about the categories of personal data %
being processed, the legal basis and purposes for processing, storage limitations, potential third parties to whom data are transferred, and details about how to get data access, rectification, or deletion. Moreover, general information about the data controller's representative and data protection officer has to be accessible. %
This information must be provided \enquote{in a concise, transparent, intelligible and easily accessible form}. %
The CCPA and other recent (draft) regulations define similar information obligations. Currently, such information is typically provided in privacy policies written in complex, ambiguous, and legalese language.

Going beyond such traditional privacy policies, several concrete designs of privacy icons have been proposed over the last decade. For instance, some have been evaluated in e-commerce scenarios \cite{holtz2010towards}, others have been proposed for online social networks \cite{iannella2010privacy}, notice \& choice mechanisms have been discussed together with visualizations \cite{cranor2012necessary}, and, recently, modular and ontology-based icons have been presented \cite{rossi2020}. Furthermore, visual nudges to encourage user interaction have been discussed extensively \cite{grafenstein2021effective}. %
Notably, such advanced transparency interfaces are not limited to privacy icons. Schaub~et~al. proposed a privacy notice design space featuring different timing, channel, modality, and control options \cite{schaub2015}. For instance, the Polisis project demonstrated how information can be automatically extracted from privacy policies and be presented through different channels \cite{hamza2018}. All of these are meant to not only comply with data protection regulations, but also to increase the trust relationship between a data controller and a data subject. Hence, data controllers have an incentive for adequate representation of their regulatory alignment and beyond.
Nevertheless, controllers are %
often lacking the expertise to implement such measures on their own apart from general guidelines (e.g., from supervisory authorities, such as the EDPB). Therefore, we aim to provide an actionable model to implement transparency-enhancing interfaces in practice.

From a technical perspective, we identify several machine-readable privacy policy languages. The most prominent approach is the Platform for Privacy Preferences Project (P3P). However, it is limited in its expressiveness and therefore not suited to meet legislative givens, which was one of the reasons it was discontinued \cite{cranor2003p3p}. Later studies, for instance as in \cite{leicht2019survey}, provide an overview and compare the different languages available. Of particular practical relevance are policy languages and accompanying developer toolkits, such as LPL (\textit{Layered Privacy Language}) \cite{gerl2018lpl}, TIRA (\textit{Transparency in RESTful Architectures}) \cite{grunewald2021tira}, and TILT (\textit{Transparency Information Language and Toolkit}) \cite{tilt}, that can capture transparency information from modern, constantly evolving, and inherently complex distributed systems %
\cite{devprivops}. Moreover, consent or preference languages, such as YaPPL (\textit{YaPPL is a Privacy Preference Language}) \cite{UlbrichtYaPPLLightweightPrivacy2018a}, GPC (\textit{Global Privacy Control})\footnote{\url{https://globalprivacycontrol.org/}}, and ADPC (\textit{Advanced Data Protection Control}) \cite{epubwu8280}, or controversially discussed industry-driven consent management platforms (e.g., IAB Europe Transparency and Consent Framework) \cite{hils2021privacy} are complementary to these, albeit with a different focus on providing legitimacy for processing instead of transparency as the main goal. %
Until now, a wide adoption of these can not be observed, since many of these technologies lack a holistic view on the privacy interaction. Consequently, we argue for a more informed approach to build upon, which is based on the Universal Design approach.  

In its first conceptualization, back in 1985, Universal Design referred mainly to issues related to usability that the disabled community experienced, especially in regard to architecture and access to buildings \cite{mace1985universal}. Gradually, the concept and principles of Universal Design begun to be adopted by different fields and areas of human experience, such as technology, education, policy and more. For example, \cite{fuchs_hci_2010} elaborates on the integration of Universal Design in human-computer interaction. In 2006, references to Universal Design were included in the Convention on the Rights of Persons with Disabilities (CRDP) of the United Nations, requiring the ratifying states to \enquote{undertake or promote research and development of universally designed goods, services, equipment and facilities [\ldots] to promote their availability and use} (Preamble). The Convention additionally provides a proper definition of Universal Design, as “the design of products, environments, programs and services to be usable by all people, to the greatest extent possible, without the need for adaptation or specialized design” (Art. 2). It is worth noticing that since the CRPD has been ratified by nearly all the members of the UN, its definition of Universal Design holds global relevance. 

The CRPD highlights the need for designs to be usable by \enquote{all people}. Scholars have interpreted it as an unambiguous reference to human diversity and the complexity of human experience, particularly in regard to access and use of ICT \cite{imrie2012universalism, ostroff2011universal}.  In practice, adopting a Universal Design approach that takes into account such diversity of needs in the designing phase of a product ensures its usability as well as safeguarding equality and non-discrimination in the process \cite{giannoumis2019conceptualizing}.

\section{General model for providing transparency information} \label{sec:model}

In this work, we will be referring to Giannoumis~and~Stein’s conceptualization of Universal Design for the information society \cite{giannoumis2019conceptualizing}. %
A set of four principles is identified with the purpose of shifting the focus from Universal Design as an outcome to Universal Design as a process: 

\begin{itemize}
    \item \textbf{Social Equality}: Equality and non-discrimination should explicitly be focal points in the design of both policy and practice, ensuring democratic access to any solution proposed. In this context, transparency in privacy interfaces and policies is necessary in order to promote social equality in the fruition of ICT. 
    \item \textbf{Human Diversity}: The complexity of the human experience(s) implies a variety of barriers that can be experienced by technology users in the creation of any product. Taking into account this aspect in addressing transparency in privacy interfaces requires the consideration of versatile interfaces to cater for the diverse needs. 
    \item \textbf{Usability and Accessibility}: It is not enough for a design to be usable if a vulnerable demographic cannot access it, and vice versa. Traditional privacy policies are often difficult to locate for non-expert data subjects, and the language they adopt is extremely specific to the legal field. Hence, advanced user interface technologies have to be considered. 
    \item \textbf{Participatory Processes}: User participation and the implementation of user feedbacks and testing are essential elements for the design of inclusive solutions. Privacy interfaces, thus, also need to include respective preference communication mechanisms (during the development and the operation). 
\end{itemize}

\begin{figure}[!ht]
\centering
\includegraphics[width=1.0\textwidth]{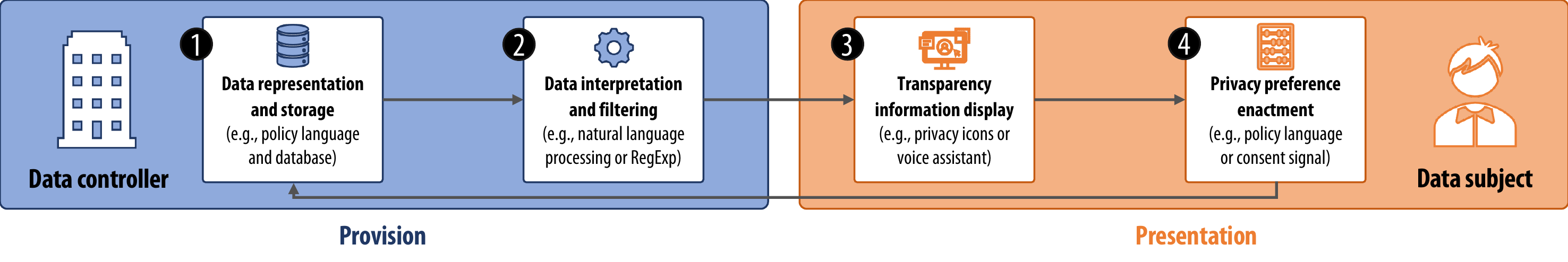}
\caption{Model of exchanging transparency information between a data controller and a data subject.}
\label{fig:architecture}
\end{figure}

Based on these considerations relating to Universal Design, we construct a general reference model for providing transparency information between a data controller and a data subject that brings together the so far isolated aspects listed above. We argue, the model of this privacy interaction explicitly needs to recognize the above-mentioned principles. We first derive our model construction considerations, depicted in Fig.~\ref{fig:architecture}, and continue explaining the key ideas.

\textbf{Distinguishing presentation from provision.} Addressing above-mentioned principles in the domain of privacy-related transparency information requires presenting transparency information in different forms and modalities, consciously tailored to the capabilities and needs of different audiences. Visually impaired users may, for instance, need high-contrast or %
acoustic interfaces, certain groups may require particularly simple language, and so forth. At the same time, the underlying transparency information to be presented does \emph{not} differ between these cases. The required form may change, while the content does not.

In a first step, we therefore consciously distinguish between the \emph{presentation} of transparency information and their originary \emph{provision}. Of these, provision happens before the presentation and comprises all activities and technical means involved in determining, compiling, and making accessible the information to be presented (the content), independently of the possibly user-specific interface (the form). The subsequent step of \emph{presenting} this information, in turn, includes all aspects related to (possibly different, needs-adapted) user interfaces, user interaction, et cetera. 

These two main stages can be assigned to different control spheres: Provision commonly happens on the side of the data controller who \enquote{determines the purposes and means of the processing}.\footnote{Responsibilities of potentially existing data processors are subsumed under the liability of the data controller which they are acting on behalf (Art.~29~GDPR).} 
In web-based scenarios, this control sphere is often equivalent to the service provider. Presentation, in turn, happens in the control sphere of the data subject (the user), employing the presentation and interaction capabilities of the client device and/or client software (e.g., web browser, operating system).
Across these two main stages and control spheres, we identify four substages of interaction required for the provision of transparency information in line with the above-mentioned principles of Universal Design. Of these, the %
first two %
belong to the main stage of \textit{provision}, %
while the latter two constitute the one of \textit{presentation}:%

\pointer{1} \textbf{Data representation and storage.} Foremost, to allow for any interaction at all, transparency-related information should be represented in a well-defined, machine-readable format. 
Such a representation %
can then be stored in potentially multiple versions, e.g., in a public database. Equally important 
is the availability of an operational and reliable API or query language to enable interoperability with other services \cite{tilt}. We emphasize this structural representation in related work on policy languages, as indicated in section~\ref{sec:related-work}. The data curation can be carried out by the controller itself, a trusted third party, or even be crowdsourced in a public repository. Having these data collected in a reasonable quality enables all further stages and versatile privacy interfaces, which outweighs the initial collection overhead.

\pointer{2} \textbf{Data interpretation and filtering.} Once the information exists in machine-readable and structured format, and as soon as it can be accessed programmatically, we can start data interpretation and filtering. Some policy languages even include certain pre-defined data transformations, such as aggregation functions.
Depending on the data representation, these processes may comprise tasks such as pre-processing (format conversion), vocabulary matching with existing privacy-related terminology for standardized wording \cite{pandit2019dpv}, or more complex natural language processing for translation or named-entity recognition for detecting cross-service data sharing. %
The necessary operations naturally depend on the data representation format and the compatibility of certain language features. Related work shows the general feasibility of doing so with multiple examples \cite{becher2022contra}. 
When transparency information for multiple data controllers is present (e.g., harvested from different APIs or taken from a public corpus), even complex sharing network analysis can be performed \cite{tilt}. In our model, we consciously differentiate between the raw storage and the actual interpretation. The model is constructed having in mind the applicability to as many use-case scenarios as possible. Since data controllers may have different intentions on what they want to present and data subjects cultivate different information needs (as pointed out in the human diversity principle of Universal Design), this step allows for much individualization as opposed to preset solutions. 

\pointer{3} \textbf{Transparency information display.} Next, after successful interpretation and filtering,
the transparency information can be presented to the data subject. Here, we emphasize the privacy notice design space \cite{schaub2015} again and underline our aim of versatile privacy interfaces as well as legal expressiveness, machine-readability, and Universal Design principles. Of these, machine-readability not only serves %
the accessibility dimension, but also the freedom of choice of client-side privacy agents. %
The display happens in the control sphere of the data subject, which opts for their desired presentation option. Notably, prior work pointed out the urgent need for adequate presentation options for recent technological advances, such as IoT devices processing personal data \cite[e.g.,][]{davies2016privacy, perez2018review}. We point out, that multiple display options can and should be offered based on the same structured transparency information. For visually impaired people, for instance, an audio or haptic interface might be considered, while others might prefer textual or visual presentations (see also \cite{zimmermann2015enhancing}).  

\pointer{4} \textbf{Privacy preference enactment.} In cases in which the subject has the option to provide privacy preferences (level of detail, legal background, technological safeguards etc.), the user interface must be capable of signaling the desired state of presentation. That is why a communication protocol between the user agent (e.g., the browser) and the service provider's infrastructure is necessary. We denote different approaches to client- and server-initiated correspondence with associated communication challenges. Furthermore, several competing standards differ in privacy signal contents, interpretation, communication, or contextual factors as pointed out by Human~et~al. \cite{human2022data}. Data controllers implementing any of these protocols usually propagate and store preferences in their infrastructure. %
As a result, the enactment of privacy preferences (guided by the informational self-determination principle) allows for individualized interfaces meeting the Universal design principles, and providing effective transparency.  

Consequently, each of the stages mentioned above in the proposed model comes with legal, technical, and possibly organizational challenges. Therefore, we hereby aim to structure the discussion about concrete system designs and to deliberately separate the different concerns. %

\section{Implementation}\label{sec:implementation}
After having explained the key concepts of our reference model, we now instantiate it by providing two implementations to %
illustrate its practical applicability and relevance in different contexts. 

\subsection{Layered Privacy Dashboard including Privacy Icons} \label{sec:privacy-icons}

In the light of well-known dysfunctionalities of textual, legalese, and all too often incomprehensible privacy policies (including transparency-related statements), it is noteworthy that the GDPR already foresees a different, more intelligible modality for providing transparency information: 
Art.~12(7f) GDPR explicitly states that \enquote{the information to be provided to data subjects [\ldots] may be provided in combination with standardized icons to give in an easily visible, intelligible and clearly legible manner a meaningful overview of the intended processing}. %
Following that, Grafenstein et al. give an overview of the visual components pertinent to the realization of a layered privacy dashboard including privacy icons, exemplarily designed for the domain of cookies \cite{grafenstein2021effective}. The study rests on participatory user workshops (in line with the Universal Design approach) that determine the information that has to be presented in order to be most effective and useful. They develop a three-layered approach in which increasing specificity and intricacies of information are presented to allow for various user preferences and competences \cite{grafenstein2021effective}, which addresses the above-mentioned Universal Design goal of human diversity. In principle, this approach is also eligible to convey other transparency information than those related to cookies, like e.g. the categories of personal data being collected or purposes of their processing.

Within this prototypical implementation, transparency is achieved through the visual components of a banner consciously designed to serve above-mentioned, participatory gathered information needs. Following a \enquote{layered approach} \cite{wp29transparency}, the first layer displays a basic icon symbolizing the availability of further information (see Fig.~\ref{subfig:privicons_1}), also facilitating usability.
The first layer can be seen as an entry point, indicating that additional transparency information is available.
In the second layer, icons that represent purposes and data categories specified in the privacy policy of the data controller are displayed. For example, Fig.~\ref{subfig:privicons_2} shows the second layer of a controller that processes data for the purpose of improving the website.
The third level acts as a comprehensive dashboard that allows the user to interact with the privacy settings and make more granular decisions (see Fig.~\ref{subfig:privicons_3}). It consists of multiple pages that reveal, for instance, information about third-party sharing and third-country transfers outside the European Union.
The information is illustrated with a network diagram that allows us to intuitively understand how information is passed on to other entities. %

Since the proposal \cite{grafenstein2021effective} was of a purely conceptual nature, our contribution lies in the actual development of a technological underpinning and its concrete implementation.
In their study, Grafenstein~et~al. already determined the effectiveness of their designs based on their evaluations. Consequently, the effectiveness evaluation of the visual representation does \textit{not} need to be repeated in this work. However, for the approach to be considered a state-of-the-art implementation according to Art.~25~GDPR, we argue that \textit{all} stages from our model need to be addressed. So far, their work focused mainly on the \textit{transparency information display} step, while the data controller's control sphere (\textit{provision} stages) was not discussed technology-wise.

\input{img/privicons_new}

Thus, we have developed a working prototype that allows users to visually access and control which data are shared with whom, for which purpose, etc. We cover all relevant steps from our proposed model and incorporated the findings from the original study \cite{grafenstein2021effective}. To do so, we applied our above-mentioned general model as follows:

\pointer{1} The underlying transparency information is represented using the \textit{Transparency Information Language and Toolkit (TILT)} \cite{tilt}, which has the required, GDPR-aligned expressiveness. %
We assume the transparency information according to Art. 12--15~GDPR to be provided by the data controller. Moreover, by querying an instance of \textit{tilt-hub}\footnote{
\url{https://github.com/Transparency-Information-Language/tilt-hub}},
which is a document store that provides several query-capable web APIs, we also have access to the transparency information for potential secondary recipients (onward transfers), as well as an overview of related third parties. %
Within this setup, TILT is used as the fundamental data representation stage, as presented in our above-mentioned model.

\pointer{2} Next, data interpretation and filtering are performed within our prototype in the control sphere of the data controller. In particular, we encoded several interpretation rules (e.g., \texttt{if-then-else} statements) that lead to the generation and subsequent display of suitable privacy icons. In fact, the provided transparency information is scanned for key terms that indicate respective risks for the data subject. For now, we interpreted different categories of personal data disclosed or the purpose specification given. However, we call for contributions to provide even more sensible content-related interpretations and summaries (e.g., as in privacy nutrition labels \cite{kelley2009nutrition}).

\pointer{3} Afterwards, all information is provided on the controller side, retrieved on the data subject side, and visually displayed on stacked layers following the underlying design proposal \cite{grafenstein2021effective} for presentation. To remove barriers to adoption, the complete dashboard and banner components are technically encapsulated and based on standard web-development technologies. They are all accessible to any programmer in an open-source code repository.\footnote{
\url{https://github.com/DaSKITA/privacy-icons}} 
Consequently, the banner does not need to be coded anew but can be easily integrated into existing websites. Hence, we minimize the cost of implementation and integration to align with the privacy by design and by default principle (Art.~25~GDPR).  %

\pointer{4} In addition, data subjects can choose different options to switch off transfers to selected third parties or by groups such as data controllers residing outside the EU. Presentation preferences are stored locally on the client device to be accessible, when revisiting the service.\footnote{Furthermore, by incorporating the YaPPL privacy preference language \cite{UlbrichtYaPPLLightweightPrivacy2018a}, we can communicate a respective signal back to the controller if desired.} %

\subsection{Interactive Privacy Chatbot and Voice Assistant through Conversational AI}\label{sec:chatbot}

Another approach to %
making transparency information more accessible for data subjects with different needs, is the usage of conversational AI \cite{harkous2016}.
To this end, we developed a \textit{Transparency Information Bot (TIBO)} %
that can be accessed in the form of a textual %
chatbot or as a speech-based virtual assistant (VA). Again, we provide our complete implementation as open-source software.\footnote{%
\url{https://github.com/DaSKITA/chatbot}}
In the following, we discuss our design choices according to the general model.

\begin{figure}[h]
 \includegraphics[width=\linewidth]{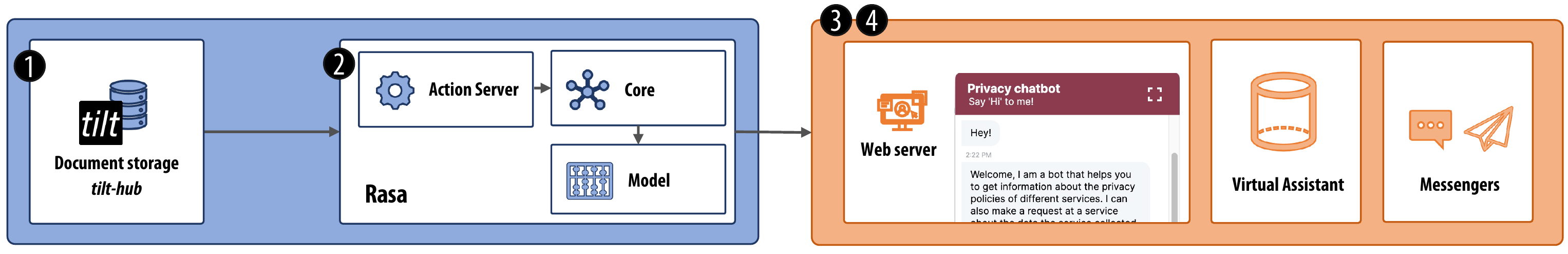}
    \caption{Architecture of the Interactive Privacy Interface using Conversational AI.
    \label{subfig:chatbot_architecture}}
\end{figure}

\pointer{1} For this prototype, we assume transparency information to be provided in the TILT ecosystem again, as described in the first implementation. Besides illustrating the advantage of different presentations being implementable on top of the same information provision, this allows for constant extension and potentially crowdsourcing of transparency information for numerous data controllers. We successfully began such efforts in a public corpus featuring dozens of real-world online services.\footnote{
\url{https://github.com/Transparency-Information-Language/tilt-corpus}}

\pointer{2} To interpret the transparency information, we implemented a %
cloud-based infrastructure at the core of TIBO (see Fig.~\ref{subfig:chatbot_architecture}), which is based on the commonly-used Rasa~X framework for conversational AI \cite{rasa2017}.
In particular, there are two core modules in the framework.
First, a natural language understanding module (Rasa NLU) receives a message from the data subject and interprets it using NLP and machine learning techniques.
Second, Rasa Core uses the interpreted data as input and computes an action (typically a response message) based on the previous conversation history, 
employing a trained classification model and various dialogue flows.
For that, a classification model is continuously trained to choose the best response for a given input from a predefined list of actions. 
For instance, we provide different dialogue flows depending on the input question. %

\pointer{3} For the display, we implemented both a web server and a frontend that features the TIBO chatbot and a skill for the popular Amazon Alexa platform in data subject's control sphere. The chatbot can, similar to the privacy dashboard, be included in existing websites to support conversational privacy dialogues.  
Through the speech-based interface, TIBO can also be used by visually-impaired people or on devices without screens (e.g., IoT devices). Researching further similar accessibility improvements -- guided by the proposed model foundations -- is considered impactful future work. %

\pointer{4} %
Additionally, the prototype enables data subjects to specify their desired preferences of detail of the information provided. They can either ask their questions in natural language or choose from recommended conversation flows. This allows experienced data subjects to investigate the peculiarities of, e.g., data sharing networks in-depth, while others are pleased with a rough overview. Moreover, the conversational AI can continuously be trained for precisely answering recurring information requests.
Future work in this stage may %
support lightweight data subject access requests. These are then supposed to be individualized upon personal preferences, for example, with regard %
to the level of completeness.

\section{Preliminary Evaluation}\label{sec:evaluation}

To demonstrate the general feasibility of the prototypes implemented according to our model and their benefit in serving data subjects' actual informedness, both have been subjected to a preliminary evaluation. Given the significantly differing prerequisites (e.g., the layered privacy dashboard using pre-existing, participatory designed visual concepts), evaluation approaches also differ between the two prototypes.

\subsection{Layered Privacy Dashboard}
For the layered privacy dashboard, we first focus on the provision-related stages of the model: %

The data representation and storage (according to stage~\pointer{1}) challenges could be addressed by using a structured policy language. We depict an example of a (imaginary) data controller in Fig.~\ref{fig:privicons}. In fact, the prototype already supports dozens of real-world services, of which we collected transparency information in our corpus repository, vividly demonstrating the practical %
viability of our approach.

Moreover, the data interpretation and filtering (\pointer{2}) could be realized having the principle of human diversity rooted in Universal Design in mind. For instance, the automatic detection of third country transfers or the summary of purposes for the related processing activities could be automated within the prototype based on the users' specific needs. Clearly, the summary and generation of privacy icons (to be then displayed) reliefs the subjects from doing the analysis work manually. In addition, the instantiation of the general model therefore contributes to a more standardized perception of transparency information. As opposed to different information banners on each and every website, the prototype enables a quick overview of the information in an always comparable setting.  

Altogether, the privacy information banner is designed to allow users to exercise their GDPR rights to transparency.%
It is based on designs already evaluated in participatory workshops and technically implements these in a practically applicable form, following our %
general model introduced above. Therefore, we can do without additional user studies. Through our implementation efforts, we contribute a working prototype that can serve as a standardized mechanism for data controllers instead of all too often deficient \enquote{Cookie banners} that neither cover all GDPR or European ePrivacy Directive transparency requirements nor being properly evaluated beforehand.

Further experiments have to optimize transparency information display (\pointer{3}) to adequately address potential misconceptions of personal data at risk. Lastly, privacy preference and default settings have to reflect information asymmetries, power imbalances, and, again, risks associated with the processing of personal data. Discussing several options under the umbrella of our model might help. Applying stage~\pointer{4} of our reference model, we allow users to spend a minimal amount of effort to control their privacy preferences once and let the browser remember the decision. For actual integration, the payload encoded in preference requests needs to be transferred back to the controller to propagate and enforce the policy \cite{pesch2021drivers}. %

\subsection{Interactive Privacy Chatbot and Voice Assistant}
Secondly, we continue with the interactive privacy chatbot TIBO. Given that it rests upon basically similar technologies for the provision-related part, we pay particular attention to the %
presentation-related stages of the model here. %
For this, we performed a participatory user study, which was carried out as follows:

\textbf{Study design.} %
The study aims to evaluate the extent to which TIBO is useful for data subjects. In particular, we want to measure the time spent to find the desired information about the privacy-related practices of a service. The secondary goal was to obtain user feedback for the further development of the conversational AI.

\textbf{Procedure.}
We selected a variant of Unmoderated Remote Usability Testing \cite{barnum_3_2021}, where the participants have to solve various tasks given to them in an online survey, with and without the assistance of TIBO. These tasks consisted of finding out relevant information, such as the name of the data protection officer, the categories of personal data being processed, or the existence of third-party sharing. 
The participants were, based on their month of birth, divided into two groups (42 percent / 58 percent), which had to find identical transparency information for two different services, namely a major German media library service (ARD~Mediathek) and the online platform of the German Federal Ministry of Justice (former German BMJV). Group A used TIBO for the ARD~Mediathek and the traditional privacy policy of the German BMJV, while group B had to solve the same tasks the other way around.  %
The participants were free in their decision to choose whether to use the chatbot or the voice assistant.
We then measured the time taken for solving the tasks. A shorter processing time of the tasks when using one of the assistants would then indicate that the prototypes fulfill their goal. %
In addition, we conducted short interviews to discuss further use cases and potentials of TIBO.

\textbf{Participants.}
19 participants evenly distributed across gender and age groups completed the online survey. With this sample size, %
results are of limited statistical significance, but nonetheless still provide valuable insights.
Almost half of the participants were female (47 percent). The majority of participants were between 30 and 45 years old (53 percent), followed by those younger than 30 (26 percent) and between 46 and 60 (21 percent). 69 percent opted for testing the voice assistant first.
We asked the participants several questions about their prior experience and demographics, while ensuring that they stay anonymous at all times. They were briefed to be included in a scientific evaluation of the tools, and they provided consent for analyzing the given inputs. We adhered to good scientific practice and common ethical standards for such user studies. The participants were recruited from the so-called Living Lab of the CheckMyVA\footnote{\url{https://checkmyva.de/}} project.

\begin{figure}[!t]
  \centering
  \includegraphics[width=0.6\linewidth]{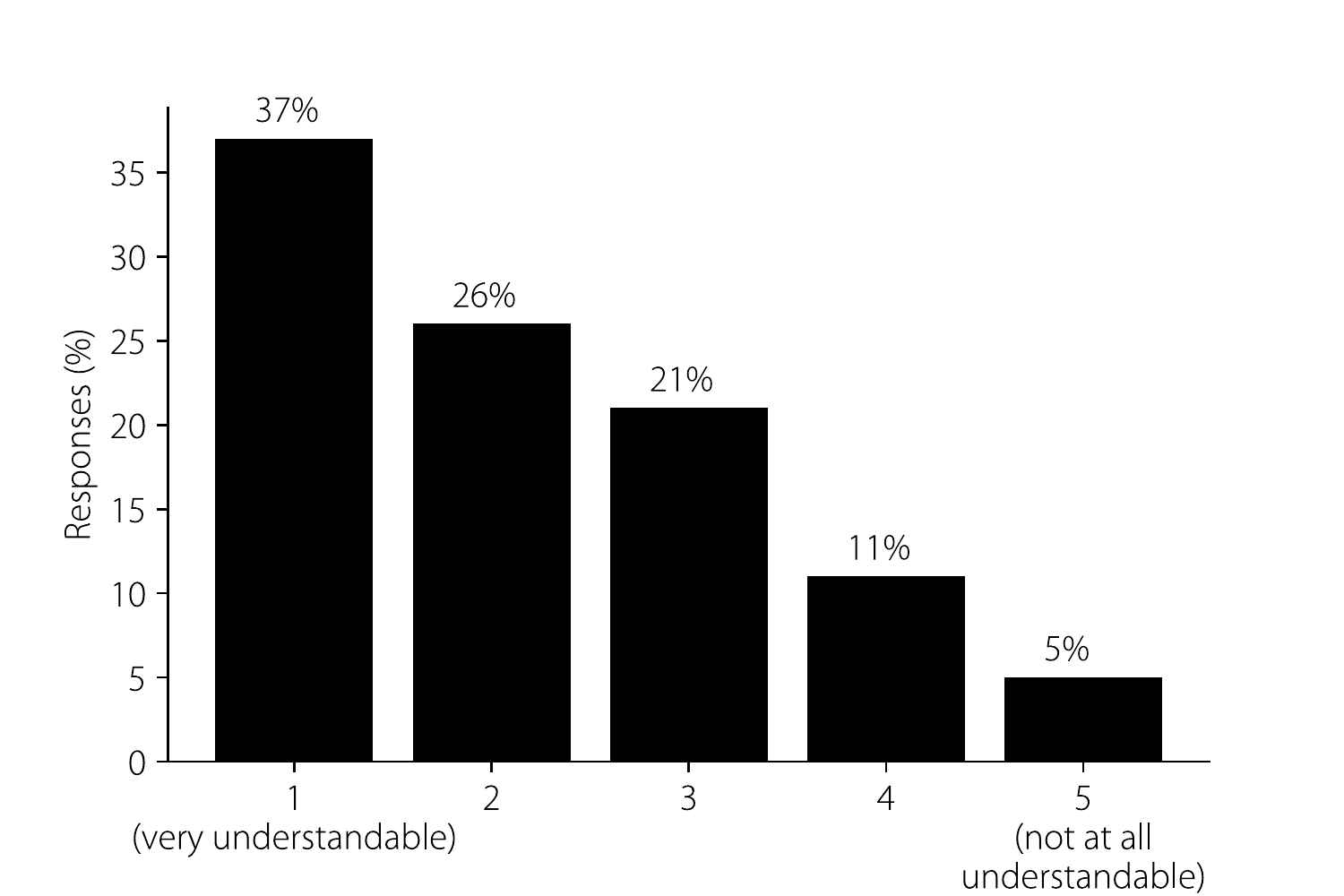}
    \caption[b]{Ease of understanding regarding responses of the chatbot and voice assistant, rated by the participants of the user study.
    \label{subfig:understandable}}
\end{figure}

\textbf{Results.} 
The results show that the use of the assistant enabled the participants to solve their tasks more quickly. The average participant spent 8:28~min to find out the required pieces of information \textit{with} the assistant, while they needed 10:51~min for the same tasks \textit{without} the assistant.
Independent of their chatbot use, they took slightly longer to complete the tasks for the BMJV than for the ARD~Mediathek (BMJV: 9:48~min; ARD~Mediathek: 9:03~min).
Participants using the chatbot were faster than those using the voice assistant, which evidently depends on the speaking rate of the virtual voice and faster processing of written text than voice. %
Considering the feedback from the participants, the answers are easy to understand (cf.~Fig.~\ref{subfig:understandable}) but there is still potential for increasing the time saved by the assistant compared to manual research, e.g., by improving response times. By and large, however, the results indicate our approach to be practically viable and valuable and to effectively help data subjects in finding and comprehending transparency information. 
One participant elaborated: \enquote{The bot makes it easier for me to find the information. It's great that the information is in the same format, allowing me to compare}.
This suggests that the value of TIBO can be increased even more if more services are included, which refers to the provision of more transparency information in stage~\pointer{1}~and~\pointer{2}.
Furthermore, the results indicate that the assistant fulfills its intended use with regard to effectiveness in display (\pointer{3}). All the participants obtained the desired information with the help of the assistant correctly and even faster than with a manual search.

Concerning the general usability, the participants noted that the chatbot does not offer the option of interrupting it during its responses. Especially during longer answers to the desired transparency information, this option should be granted in future versions. In addition, a navigation function should be implemented so that jumping between menu and selection items is possible, extending the preference enactment in stage~\pointer{4}. Overall, 53 percent of the participants stated they would use the assistant again to obtain privacy-related information, which underlines the general effectiveness of the approach. Future versions of the assistant are currently evaluated closely to the ongoing development progress.

Overall, the participants envisioned the primary usage of TIBO to check the privacy policy before they are starting to use a service. For this purpose, many of them imagine it on a website of an independent third party, e.g. a consumer center.
One participant sees the chatbot's greatest potential for children or people with learning disabilities who are not familiar with the usual privacy statements. For these people, TIBO - possibly in easy language - could be a good fit in obtaining comprehensible information. This optimally aligns with our intention of using conversational AI as a presentation means (in stages~\pointer{3} and \pointer{4}), which again addresses the universal design principles of increased usability and accessibility.

\section{Discussion, Conclusion, and Outlook}\label{sec:outlook}

We proposed a general model for the provision of transparency information while embedding the legal responsibilities and rights, technical challenges, and some cross-cutting aspects induced by diverse fields. All scholars and professionals in the field are invited to take the model as a reference for proposing related systems. Clearly, each proposed stage has to be explored in much more detail. %
In the same vein, more studies need to explore the suitable interpretation and filtering methods. These should in particular incorporate comparable transparency metrics \cite{metrics}.

Following our general model, we proposed two advanced transparency interfaces that enable context-, preference-, and competence-adaptive provision of transparency information based on machine-readable representations. To the best of our knowledge, this is the first comprehensive work on providing transparency information in a structured, machine-readable format across the control spheres of data controller and subject, presenting the information through different channels, but using the same underlying data, and guided by overarching Universal Design principles. We demonstrated the practical viability through the user study of the chatbot and the virtual assistant while incorporating the already-evaluated UI/UX design of the layered privacy dashboard. We found conversational AI makes transparency information significantly more accessible, thereby enabling users to make well-informed decisions about the processing of personal data concerning them faster than with traditional means.

Open challenges remain in large-scale integration efforts across diverse user interface technologies and platforms. We already provided two viable proof of concept implementations. Evidently, as pointed out above, the design space of an effective privacy notice is large \cite{schaub2015}. In this work, we demonstrate the actionable guidance of our reference model and aim for more versatile implementations in that space, replacing existing \enquote{dark patterns}. In addition, new and current draft regulations of transparency provisions, such as the European Data Governance Act, Digital Services Act, or AI Act, are introducing extended transparency requirements. As soon they enter into force, affected data controllers, in particular large-scale service providers, are obligated to implement comparable transparency measures. 

Future work, among others, incorporates %
participatory learning of privacy preferences and crowdsourced approaches for collecting and providing transparency information (e.g., through privacy agents). Such data could also be used for the wider prospects of advanced transparency and accountability mechanisms for practical privacy engineering and usable privacy. In all the mentioned scenarios, the distinction and logical separation of information provision and presentation is key to a sustainable technological development: System engineers can build efficient data provision formats and communication protocols, while UI/UX designers can conceptualize more usable and accessible interfaces (guided by legal expertise) together, when they agree to stick to the proposed model. This ultimately results in satisfactory components for both data controllers and subjects.%

%% file: img/privicons_new.tex
\begin{figure}[t]

\begin{minipage}{1.\textwidth}

  \centering
    \begin{subfigure}[t]{0.5\textwidth}
        \centering
        \includegraphics[width =.2\textwidth]{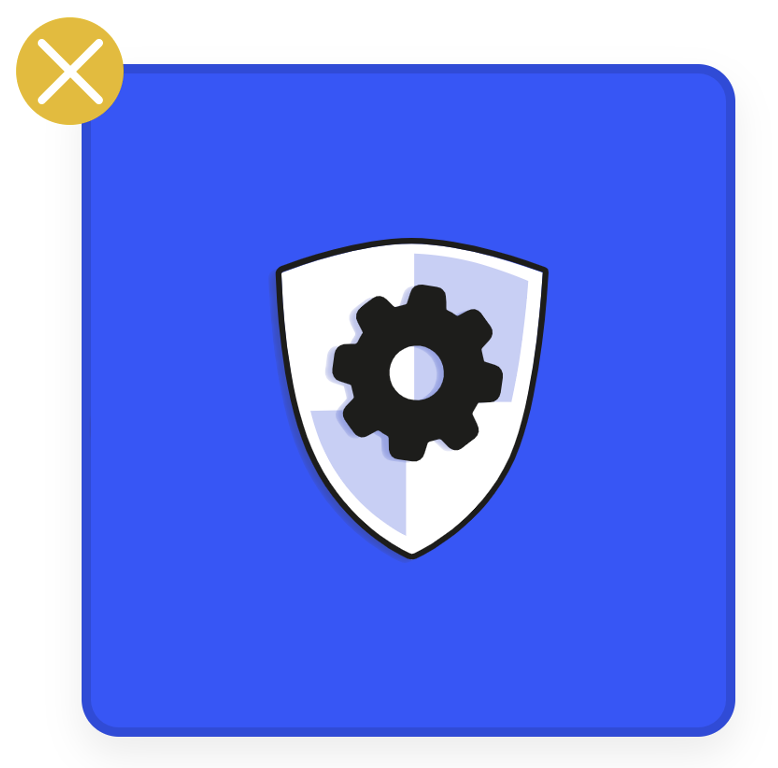}
        \caption{Level 1. Entry point. \label{subfig:privicons_1}}
    \end{subfigure}%
    ~ 
    \begin{subfigure}[t]{0.5\textwidth}
        \centering
        \centering\includegraphics[width =.55\textwidth]{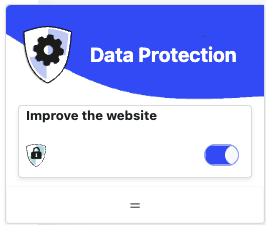}
        \caption[b]{Level 2. Purpose and risk icon. \label{subfig:privicons_2}}
    \end{subfigure}
\end{minipage}

        \vspace{0.05\textwidth} 
\begin{minipage}{1.\textwidth}
  \begin{subfigure}[c]{\textwidth}
    \centering\includegraphics[width=1\textwidth]{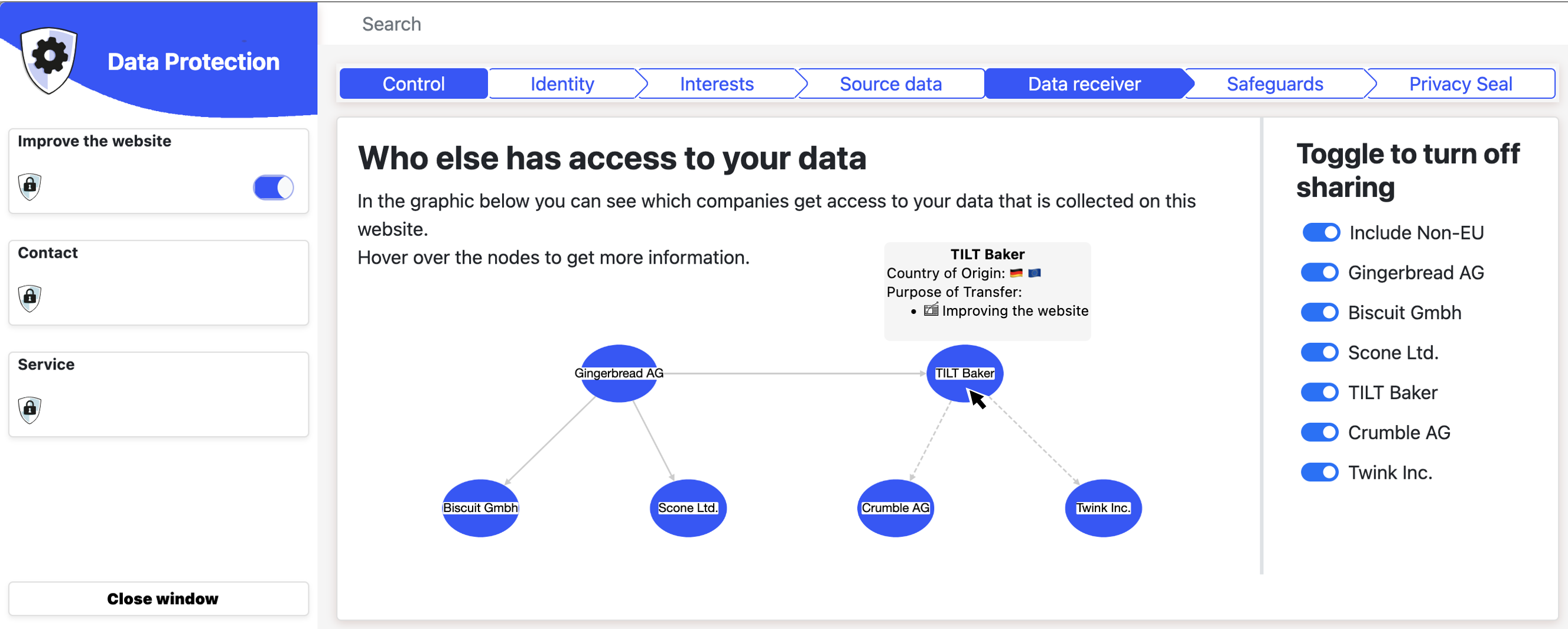}
    \caption{Level 3. Detailed view on data receivers depicting an exemplary sharing network. \label{subfig:privicons_3}}
  \end{subfigure}
\end{minipage}

\caption{Functional prototype of the layered privacy dashboard based on the design of \cite{grafenstein2021effective} and enabled using machine-readable transparency and consent information encoded in the TILT \cite{tilt} and YaPPL \cite{UlbrichtYaPPLLightweightPrivacy2018a} policy languages.}
\label{fig:privicons}
\end{figure}